\documentclass[11pt,a4paper]{article}
\usepackage[T1]{fontenc}
\usepackage[utf8x]{inputenc}
\usepackage{graphicx}
\usepackage[margin=1in]{geometry}
\usepackage{rotating}
\usepackage{multirow}
\usepackage{nicefrac}
\usepackage{tabularx}
\usepackage{placeins}
\usepackage{tabulary}
\usepackage{booktabs}
\usepackage{setspace}
\usepackage{color}
\usepackage{setspace}
\usepackage{bigstrut}
\usepackage[bottom]{footmisc}
\usepackage[colorlinks=true,urlcolor=blue,linkcolor=blue,citecolor=blue]{hyperref}
\usepackage{threeparttable}
\usepackage{dcolumn}
\newcolumntype{G}{D..{6.4}}
\usepackage{amsmath}
\usepackage{amssymb}
\usepackage{enumerate}   
\usepackage{todonotes}
\usepackage{array}
\usepackage{pdflscape}
\newcolumntype{L}[1]{>{\raggedright\let\newline\\\arraybackslash\hspace{0pt}}m{#1}}
\newcolumntype{C}[1]{>{\centering\let\newline\\\arraybackslash\hspace{0pt}}m{#1}}
\newcolumntype{R}[1]{>{\raggedleft\let\newline\\\arraybackslash\hspace{0pt}}m{#1}}
\newcolumntype{Y}{>{\centering\arraybackslash}X}

\usepackage{siunitx}
\usepackage{newtxtext,newtxmath}

\usepackage{eqparbox}

\usepackage[title,titletoc]{appendix}
\makeatletter

\renewenvironment{appendices}{%
    \begin{oldappendices}%
    \renewcommand{\thefigure}{\ifnum \c@section>\z@ \thesection.\fi\@arabic\c@figure}%
    \@addtoreset{figure}{section}%
    \renewcommand{\thetable}{\ifnum \c@section>\z@ \thesection.\fi\@arabic\c@table}%
    \@addtoreset{table}{section}%
}{%
    \end{oldappendices}%
}\makeatother

\usepackage{natbib}
\let\natbibcitet\citet
\renewcommand\citet{\bibpunct{(}{)}{,}{a}{,}{,}\natbibcitet}
\let\natbibcitep\citep
\renewcommand\citep{\bibpunct{(}{)}{;}{a}{,}{;}\natbibcitep}
\newcommand{\bi}{\begin{itemize}}
\newcommand{\ei}{\end{itemize}}
\newcommand{\be}{\begin{equation}}
\newcommand{\ee}{\end{equation}}
\defcitealias{ieo14}{IEO, 2014}

\long\def\symbolfootnote[#1]#2{\begingroup%
\def\thefootnote{\fnsymbol{footnote}}\footnote[#1]{#2}\endgroup}

\usepackage{sectsty}
\allsectionsfont{\normalsize}

\makeatletter

\@addtoreset{equation}{section}
\makeatother

\usepackage[affil-it]{authblk}

\usepackage{caption}
\usepackage{subcaption}
\makeatletter\let\p@subfigure\thefigure\makeatother

\usepackage{cleveref}
\crefname{chapter}{Chapter}{Chapters}
\crefname{section}{Section}{Sections}
\crefname{subsection}{Section}{Sections}
\crefname{subsubsection}{Section}{Sections}
\crefname{figure}{Figure}{Figures}
\crefname{table}{Table}{Tables}
\crefname{equation}{Equation}{Equations}
\crefname{appendix}{Appendix}{Appendices}

\newcolumntype{d}[1]{D{.}{.}{#1}}

\title{
	\LARGE \textbf{Modeling European regional FDI flows using a Bayesian spatial Poisson interaction model}
}

\usepackage{authblk}

\author[1]{Tam\'{a}s Krisztin\thanks{\textit{Corresponding author}: Tam\'{a}s Krisztin, International Institute for Applied Systems Analysis (IIASA), Schlossplatz 1, 2361 Laxenburg, Austria. \textit{E-mail}: \href{mailto:krisztin@iiasa.ac.at}{krisztin@iiasa.ac.at}. The research carried out in this paper was supported by funds of the Oesterreichische Nationalbank (Jubilaeumsfond project number: 18116), and of the Austrian Science Fund (FWF): ZK 35 }}
\author[2]{Philipp Piribauer}

\affil[1]{International Institute for Applied Systems Analysis (IIASA)}
\affil[2]{Austrian Institute of Economic Research (WIFO)}

\date{\vspace{-5ex}}



\begin{document}
\onehalfspacing
\graphicspath{{figs/}}
\maketitle

\begin{abstract}
\noindent
This paper presents an empirical study of spatial origin and destination effects of European regional FDI dyads. Recent regional studies primarily focus on locational determinants, but ignore bilateral origin- and intervening factors, as well as associated spatial dependence. This paper fills this gap by using observations on interregional FDI flows within a spatially augmented Poisson interaction model. We explicitly distinguish FDI activities between three different stages of the value chain. Our results provide important insights on drivers of regional FDI activities, both from origin and destination perspectives. We moreover show that spatial dependence plays a key role in both dimensions.
\\
\\
\textbf{Keywords:} Spatial interaction model, Bayesian Poisson model, regional FDI flows, European regions, spatial random effects.
\\
\\
\textbf{JEL Codes:} C11, C21, F23, R11
\end{abstract}

\newpage

\section{Introduction}


Recent decades have shown a rapid growth of worldwide foreign direct investment (FDI), which  led to increased efforts in research to understand the economic determinants of FDI activities. Classical explanations focus on the factors driving firms to become multinational. The Ownership-Localization-Internalization theory (see \citealt{dunning2001oli}) explains firms' motivation as an effort to internalize transaction costs and reap the benefits of externalities stemming from strategic assets.

A large alternative strand of empirical literature builds on trade theory. In this context the  drivers of FDI activity are the need for larger sales markets, cheaper source markets, and the willingness to reach a technological frontier (\citealt{markusen1995boundaries}). Following  empirical international economics literature, FDI flows  are usually captured within the context of a bilateral spatial interaction model framework. The main advantage of this approach is that it specifically accounts for the role of origin- and destination-specific factors, as well as intervening opportunities. For an overview on the determinants of FDI activities and  the location choice of multinationals, see \cite{basilekayam}, \cite{blonigenpiger}, or \cite{blonigen2005review}.  

Due to the scarcity of data on FDI activities on a subnational scale, the vast majority of the empirical literature focuses on country-specific FDI patterns. A subnational perspective, however, would allow for in-depth decomposition of the spatial patterns of FDI flows, since FDI sources and destinations are not uniformly distributed within a country, but tend to be spatially clustered. Multiple studies focusing on regional investment decisions of multinational companies \citep{crescenzi2013innovation,Ascani2016} emphasize within-country heterogeneity of FDI decisions, which can exceed cross-country differences. However, a major gap in the literature is that regional level studies only focus on the destination of FDI decisions, and largely neglect to account for origin-specific factors, as well as intervening opportunities in a subnational context. However, a simultaneous treatment appears particularly important for providing a complete picture on third-regional spatial interrelationships in both source- as well as destination-specific characteristics \citep{leibrecht2014}. Moreover, neglecting to take into account both origin, destination, and third region effects, can lead to biased parameter estimates \citep{baltagi2007estimating}.

The present paper aims to fill these gaps by focusing on subnational FDI flows in a European multi-regional framework and explicitly accounting for origin-, destination-, as well as third region-specific factors. In this paper we make use of subnational data from the \textit{fDi Markets} database, which reports on bilateral FDI flows, with detailed information on the source and destination city. This can be compiled to multiple dyadic format, that is each region pair appears twice, corresponding to FDI flowing from one region to the other and vice versa. A specific virtue of the database is that it distinguishes FDI flows by their respective business activity. This allows us to contrast the impact of origin, destination, and third region effects across multiple stages of the global values chain. 

When adopting a subnational perspective, it is crucial to control for spatial dependence, as its presence in regional data is well documented \citep{LeSage2009}. Even national-level empirical applications clearly document the presence of spatial spillovers on FDI activities. An influential example is the work by \cite{Blonigen2007fdi}, who analyse the determinants of US outbound FDI activities in a cross-country framework, while explicitly accounting for spatial dependence among destinations. Further studies which document the presence of spatial issues amongst bilateral (national) FDI activities include \cite{pintar2016austrian},  \cite{regelink2015spatial}, \cite{chou2011impact}, \cite{garretsen2009fdi}, \cite{poelhekke2009foreign}, or  \cite{baltagi2007estimating}.  

We therefore employ an econometric framework in the spirit of \cite{Koch2015} and \cite{LeSage2007} which captures not only third-regional effects but also spatial dependence using spatially-augmented random effects. Estimation is achieved using work by \cite{Fruhwirth-Schnatter2009}, allowing us to deal with the high-dimensional specifications in a flexible and computationally efficient way.


The remainder of the paper is organized as follows. Section 2 presents the proposed spatial interaction model, which is augmented by spatial autoregressive origin- and destination-specific random effects, intended to capture spatially dependencies, as well as so-called third region effects. Section 3 details the FDI data, the considered determinants, as well as our selection of regions. In Section 4 we assess the determinants of European interregional FDI flows across different stages of the global value chain. The analysis is performed using information on FDI dyads covering $266$ NUTS-2 regions in the period 2003 to 2011. Section 5 concludes.

\section{A spatial interaction model for subnational FDI flows}
\label{s:4_spatial_gravity}

This section presents the model specification used for the empirical analysis. It is worth noting that the spatial econometric model is similar to work by \cite{LeSage2007}, who aimed at modelling regional knowledge spillovers in Europe. An efficient Bayesian estimation approach for the employed multiplicative form of the Poisson model with spatial random effects is provided in the Appendix.\footnote{Detailed R codes for running the proposed model are available upon request.}

Let $\boldsymbol{y}$ denote an $N \times 1$ vector containing information on the number of FDI flows between $n$ regions.\footnote{It is worth noting that in the present study $N$ is of lower dimension than $n^2$, since FDI dyads by construction exhibit no own-regional and no own-country flows.} In the classic spatial interaction model framework the flows are regressed on correspondingly stacked  origin-, destination-, and distance-specific explanatory variables, as well as their spatially lagged counterparts. $\boldsymbol{X}_o$ and $\boldsymbol{X}_d$ denote $N \times p_X$ origin- and destination-specific matrices of explanatory variables, respectively.   Distances and further intervening factors between the $n$ regions are captured by the $N \times p_D$ matrix $\boldsymbol{D}$.\footnote{Detailed information on the straightforward construction of the origin- and destination-specific matrices of explanatory variables $\boldsymbol{X}_o$ and $\boldsymbol{X}_d$ from an $n\times p_X$ dimensional matrix of explanatory variables is provided in \cite{LeSage2009}. \cite{LeSage2009} also provide detailed guidelines on the convenient construction of origin- and destination-specific spatial weight matrices.} Extending the standard model specification with local spillover effects as well as spatial random effects, we consider a Poisson specification of the form:
\begin{align}
\boldsymbol{y} &\sim \mathcal{P}(\boldsymbol\lambda) \notag\\
\boldsymbol{\lambda} &= \exp\left(\alpha_0 + \boldsymbol{X}_o \boldsymbol{\beta}_o + \boldsymbol{X}_d \boldsymbol{\beta}_d +   \boldsymbol{D} \boldsymbol{\gamma}_D + \boldsymbol{W}_o \boldsymbol{X}_o \boldsymbol{\delta}_o + \boldsymbol{W}_d \boldsymbol{X}_d \boldsymbol{\delta}_d +\boldsymbol{V}_o\boldsymbol{\theta}_o + \boldsymbol{V}_d\boldsymbol{\theta}_d  \right), \label{eq:random_effects_model}
\end{align}
where $\mathcal{P}(\cdot)$ denotes the Poisson distribution and $\alpha_0$ is an intercept parameter. $\boldsymbol{\beta}_o$, $\boldsymbol{\beta}_d$, and $\boldsymbol{\gamma}_D$ are parameter vectors corresponding to $\boldsymbol{X}_o$, $\boldsymbol{X}_d$, and $\boldsymbol{D}$, respectively. The spatial lags of the covariates are captured by $\boldsymbol{W}_o \boldsymbol{X}_o$ and $\boldsymbol{W}_d \boldsymbol{X}_d$, with $\boldsymbol{\delta}_o$ and $\boldsymbol{\delta}_d$ denoting the respective $p_X \times 1$ vectors of parameters. Through these spatial lags we explicitly capture the so-called third region effects \citep{baltagi2007estimating}, that is origin- and destination-specific spillovers from neighbouring regions. Neighbourhood effects are governed by non-negative, row-stochastic spatial weight matrices, which contain information on the spatial connectivity between the regions under scrutiny. Our Poisson spatial interaction model includes separate spatial weight matrices $\boldsymbol{W}_o$ and $\boldsymbol{W}_d$ to account for origin- and destination-specific third regional effects, respectively. 

Origin-based random effects are captured by the term $\boldsymbol{V}_o\boldsymbol{\theta}_o$, where $\boldsymbol{V}_o$ denotes an $N\times n$ matrix of origin-specific dummy variables with a  corresponding $n \times 1$ vector $\boldsymbol{\theta}_o$. Similarly, the $n \times 1$ vector $\boldsymbol{\theta}_d$ captures regional effects associated with the destination regions' matrix of dummy variables $\boldsymbol{V}_d$. We follow work by \cite{LeSage2007} and introduce a further source of spatial dependence via the $n \times 1$ regional effect vectors $\boldsymbol{\theta}_o$ and $\boldsymbol{\theta}_d$, which are assumed to follow a first-order spatial autoregressive process: 
\begin{align}
\boldsymbol{\theta}_o &= \rho_o \boldsymbol{W} \boldsymbol{\theta}_o + \boldsymbol{\nu}_o \hspace{2em} \boldsymbol{\nu}_o = \mathcal{N}\left(\boldsymbol{0},\phi_o^2 \boldsymbol{I}_n \right) \label{eq:origin_fx}\\
\boldsymbol{\theta}_d &= \rho_d \boldsymbol{W} \boldsymbol{\theta}_d + \boldsymbol{\nu}_d \hspace{2em} \boldsymbol{\nu}_d = \mathcal{N}\left(\boldsymbol{0},\phi_d^2 \boldsymbol{I}_n \right), \label{eq:destination_fx}
\end{align}
where $\rho_o$ and $\rho_d$ denote origin- and destination-specific spatial autoregressive (scalar) parameters, respectively. $\boldsymbol{W}$ denotes an $n\times n$ row-stochastic spatial weight matrix with known constants and zeros on the main diagonal. 

The disturbance error vectors $\boldsymbol{\nu}_o$ and $\boldsymbol{\nu}_d$ are both assumed to be independently and identically normally distributed, with zero mean and $\phi_o^2$ and $\phi_d^2$ variance, respectively. Note that this assumption implies a one-to-one mapping to origin- and destination-specific normally distributed random effects in the case of $\rho_o = 0$ and $\rho_d = 0$. For a row-stochastic $\boldsymbol{W}$, a sufficient stability condition may be employed by assuming the spatial autoregressive parameters $\rho_o$ and $\rho_d$ to lie in the interval $-1 < \rho_o, \rho_d < 1$ (see, for example, \citealt{LeSage2009}).

\section{Bilateral FDI data and regions}

Our data set comprises observations on regional FDI dyads for 266 European NUTS-2 regions in the period 2003 to 2011. A complete list of the regions in our sample is provided in Table \ref{tab:Appendix_regions} in the Appendix.

Observations on regional cross-border greenfield FDI investments stem from the \textit{fDi Markets} database. This database is maintained by fDi Intelligence, which is a specialist division of the Financial Times Ltd. The  provided data draws on media and corporate sources to report on the sources and hosts of FDI flows (detailed by country, region, and city), industry classifications, as well as the level of capital investment. \cite{crescenzi2013innovation} report several robustness tests and detailed comparisons with official data sources. They confirm the reliability of the \textit{fDi Markets} data set, especially with regard to the reported spatial distribution of FDI investments.

Our dependent variables are based on the total amount of inflows from European regions in the period 2003 to 2011. Since  the \textit{fDi Markets} data base also contains information on several distinct business activities for both origin and host companies, we follow previous studies  by \cite{Ascani2016} and study the determinants of regional FDI dyads at different stages of the value chain.  This information is valuable as investor companies maximize their utility with respect to their position along the value chain. Since specifics of the investor company, as well as details on the FDI investment are largely unobserved, it is crucial to account for the heterogeneity in investor decisions by subdividing industry activities relative to their position along the value chain  (see, for example, \citealt{Ascani2016}). We therefore define three different classifications: \textit{Upstream}, \textit{Downstream}, and \textit{Production}. The classification adopted in this paper builds on general classifications of the value chain by \cite{Sturgeon2008} and closely tracks the ones employed by \cite{crescenzi2013innovation} and \cite{Ascani2016}.

Specifically, the upstream category comprises conceptual product development including design and testing, as well as management and business administration activities. The downstream category summarizes consumer-related activities such as sales, product delivery, or support. Finally, the production category includes activities related to physical product creation, including extraction, manufacturing, as well as recycling activities. A complete list of the employed global value chain classification is provided in Table \ref{tab:classification} in the Appendix.

Our choices for explanatory variables are motivated by  recent literature on (regional) FDI flows as well as regional growth empirics (see, for example, \citealt{cuaresma2018human}, \citealt{blonigenpiger}, \citealt{leibrecht2014}, or \citealt{blonigen2005review}). In most  gravity-type models a region's ability to emit and attract FDI flows is chiefly captured by its economic characteristics. Our main indicator for economic characteristics is the regions' market size, proxied by regional gross value added. To control for the degree of urbanization both in origin and host regions we also include regional population densities as an additional covariate. Empirical evidence suggests \citep{coughlin1991state,huberfischerpiribauer}  that higher wages have a deterrent effect on investment. We proxy this in our model by including the average compensation of employees per hour worked as an explanatory variable.

We account for the regional industry mix by including the share of employment in manufacturing and construction (NACE classifications B to F), as well as services (NACE G to U). We moreover include typical supply-side quantities such as regional endowments of human and knowledge capital. To proxy regional human capital endowments we include two different variables. The first variable measures regional tertiary education attainment shares labelled higher education workers. A second variable labelled lower education workers is proxied by the share of the working age population with lower secondary education levels or less.

We use data on patent numbers to proxy regional knowledge capital endowments. Patent data exhibit particularly desirable characteristics for this purpose, since they can be viewed as a direct result of research and development activities (\citealt{fischerlesage_knowledge}). In order to construct regional knowledge stocks we use the perpetual inventory method. We follow \cite{fischerlesage_convergence} and \cite{fischerlesage_knowledge} to construct knowledge capital stocks $K_{it}$ for region $i$ in period $t$. Specifically, we define $K_{it}=(1-r_K)K_{it-1}+P_{it}$, where $r_K=0.10$ denotes a constant depreciation rate and $P_{it}$ denotes the number of patent applications in region $i$ at time $t$.

The matrix $\boldsymbol{D}$ includes several different distance metrics. First and foremost, we include the geodesic distance between parent and host regions. Recent empirical literature also consider common language as a potential quantity in $\boldsymbol{D}$ (see \citealt{Krisztin2015}, or \citealt{blonigenpiger}). We measure whether the same official language is present in the source and host regions through a dummy variable. Information on official national and minority languages is obtained from the \textit{European Commission}.

Several studies on FDI flows also highlight the importance of corporate tax rates as a potential key quantity to attract FDI inflows (see \citealt{blonigenpiger}, \citealt{leibrecht2014}, and \citealt{Bellak2009}). Lower corporate income tax rates in the host region as compared to the origin region are thus expected to increase the potential attractiveness of FDI inflows. Matrix $\boldsymbol{D}$ therefore also contains the (country-specific) difference in corporate income tax rates between origin and destination regions. Larger differences are expected to be associated with increasing FDI inflows. 

In order to alleviate potential endogeneity problems, we moreover measure all explanatory variables at the beginning of our sample (that is in 2003).\footnote{To assess the robustness of the results we also estimated a model where the explanatory variables were averages from 2003 to 2011. Overall the estimated quantities and their statistical significance remained unchanged.} For specification of the spatial weight matrix  we rely on a row-stochastic seven nearest neighbour specification.\footnote{A series of tests using different number of nearest neighbours  for the neighbourhood structure appeared to affect the results in a negligible way.} Data on the variables used stem from the \textit{fDi Markets}, \textit{Cambridge Econometrics}, as well as the \textit{Eurostat} regional databases. Detailed information on the construction of the dependent and explanatory variables used are presented in Table \ref{tab:variables}. 

\begin{table*}[h!tbp]
\caption{Variables used in the empirical illustration}\vspace*{-1.8em}
\small
\begin{center}
\begin{threeparttable}
\begin{tabular*}{\textwidth}{@{\extracolsep{\fill}} >{\raggedright}p{0.05\textwidth}>{\raggedright\arraybackslash}p{0.25\textwidth}>{}p{0.65\textwidth}}
\toprule
& \textbf{Variable} & \textbf{Description}\\
\midrule
\parbox[t]{2mm}{\multirow{4}{*}{$\boldsymbol{y}$}} & Upstream & FDI inflows associated with upstream activities. \textit{Source:} \textit{fDi Markets} \bigstrut \\
& Downstream & FDI inflows associated with downstream activities. \textit{Source:} \textit{fDi Markets} \bigstrut\\
& Production & FDI inflows associated with production activities. \textit{Source:} \textit{fDi Markets} \bigstrut \\
\midrule

\parbox[t]{2mm}{\multirow{17}{*}{$\boldsymbol{X}$}} & Market size & Proxied by means of regional gross value added, in log terms. \textit{Source: Cambridge Econometrics} \bigstrut \\

& Population density & Population per square km, in log terms. \textit{Source: Cambridge Econometrics} \bigstrut \\

& Compensation per hour & Compensation of employees per hours worked, in log terms. \textit{Source: Cambridge Econometrics} \bigstrut \\

& Employment in industry & Share of NACE B to F (industry and construction) in total employment. \textit{Source: Cambridge Econometrics} \bigstrut \\

& Employment in services & Share of NACE G to U (services) in total employment. \textit{Source: Cambridge Econometrics} \bigstrut \\

& Lower education workers & Share of population (aged 25 and over) with lower education (ISCED levels 0-2). \textit{Source: Eurostat} \bigstrut \\

& Higher education workers& Share of population (aged 25 and over)  with higher education (ISCED levels 6+). \textit{Source: Eurostat} \bigstrut \\

& Regional knowledge capital & Knowledge stock formation measured in terms of patent accumulation, in log terms. \textit{Source: Eurostat} \bigstrut \\  
\midrule
\parbox[t]{2mm}{\multirow{6}{*}{$\boldsymbol{D}$}}& Geographic distance & Geodesic distance between source and host region. \textit{Source: Eurostat} \bigstrut \\
& Difference in tax rates & Country-specific top statutory corporate income tax rates (including surcharges). Measured by means of difference between source and host region. \textit{Source: Eurostat} \bigstrut \\
& Common language & Dummy variable, 1 denotes that the regions share a common official language, 0 otherwise. \textit{Source: European Commission}  \bigstrut \\

\bottomrule
\end{tabular*}
\begin{tablenotes}[para,flushleft]
\footnotesize{\textit{Notes}: ISCED and NACE refer to the international standard classification of education and the second revision of the statistical classification of economic activities in the European community, respectively. 
}
\end{tablenotes}
\end{threeparttable}
\end{center}
\label{tab:variables}
\end{table*}

\FloatBarrier

\section{Empirical results}

This subsection presents the empirical results obtained from 15,000 posterior draws after discarding the first 10,000 as burn-ins. Running multiple chains with alternating starting values did not affect the empirical results, which also provides evidence for sampler convergence.

Posterior quantities for upstream-, downstream-, and production-related investment flows are presented in Tables \ref{tab:upstream}, \ref{tab:downstream}, and \ref{tab:production}, respectively. Each table reports posterior means and posterior standard deviations for the quantities of interest. Statistical significance of the respective posterior mean estimates is based on a 90\% credible interval and highlighted in bold. The first block in each table presents origin- and destination-specific slope parameter estimates, respectively. These estimates are reported for both own region characteristics as well as their spatial lags or third region characteristics (\citealt{baltagi2007estimating}). In the spatial econometrics literature, the former are often referred to as average direct impacts. The  spatially lagged counterparts denote average indirect (or spillover) impacts (\citealt{LeSage2009}). The second block in each table reports posterior summary metrics for the spatial autoregressive origin and destination random effects. The third and last block in each table shows posterior inference for the variables used in the distance matrix $\boldsymbol{D}$.


\subsection*{Origin- and destination-specific core variables}

Table \ref{tab:upstream} reports posterior parameter estimates for upstream FDI (most notably consisting of business services and headquarters). Starting with the key drivers for regions producing FDI outflows in upstream-related activities, Table \ref{tab:upstream} shows particularly strong evidence for the importance of the own-regional \textit{market size} and \textit{population density}. In addition, the corresponding third-regional effects are significant and negative. For example, an increase in the market size restricted only to neighbouring regions thus decreases the amount of FDI outflows from a given region. The table also suggests a particularly accentuated importance of a well educated working age population (\textit{higher education workers}) in the origin region. The estimated impact appears much more pronounced as compared to downstream and production FDI. Moreover, for upstream FDI the third region effect associated with the \textit{higher education workers} variable also appears to be positive and highly significant. Own-regional \textit{knowledge capital} endowments appear to be positively associated with the generation of upstream FDI outflows. However, the impacts of \textit{regional knowledge capital} endowments for upstream FDI outflows appear rather muted as compared to the other types of FDI considered. Interestingly, Table \ref{tab:upstream} shows negative third-regional impacts for \textit{knowledge capital}.  Unlike other types of FDI under scrutiny, the \textit{compensation per hour} variable only appears to have a significant impact for own-regional upstream FDI outflows. 

Inspection of the regional determinants to attract upstream FDI inflows shows some interesting similarities to the origin-specific characteristics. This holds particularly true for the \textit{market size} and \textit{population density} variables. Both destination-specific variables show a positive and highly significant own-regional impact, with negative (and significant) spatial lags. Similar to the origin specific determinants of upstream FDI, the corresponding host-specific impacts appear more pronounced as in other activity types. This finding is in line with \cite{henderson2008manufacturing}, \cite{defever2006functional}, or \cite{duranton2005sectoral}, who highlight that the location choice of business services and headquarters related activities are particularly driven by functional aspects (rather than by sectoral aspects) and typically tend to be located in urban agglomerations. Regional FDI inflows associated with upstream investment activities moreover appear to be particularly attracted by regions with a higher specialization in the services sector (\textit{employment in services}), relative to the agriculture sector (which serves as the benchmark in the specifications).

\begin{table}[h!tb] 
\centering
\begin{threeparttable}[b]
  \centering
  \caption{Posterior parameter estimates  for  FDI associated with upstream value chains. \label{tab:upstream}}
  \small
\begin{tabularx}{.9\linewidth}{@{}l c*{4}{Y}@{} }
\toprule
\multicolumn{1}{c}{\multirow{2}[2]{*}{Variable}} & \multicolumn{2}{c}{Origin} & \multicolumn{2}{c}{Destination} \\
   & \multicolumn{1}{c}{Mean} & \multicolumn{1}{c}{Std. Dev.} & \multicolumn{1}{c}{Mean} & \multicolumn{1}{c}{Std. Dev.} \\
\midrule
Market size & \textbf{1.26} & 0.12 & \textbf{1.32} & 0.07 \\
Population density & \textbf{0.33} & 0.12 & \textbf{0.29} & 0.04 \\
Compensation per hour & \textbf{-0.59} & 0.31 & \textbf{-0.94} & 0.17 \\
Employment in industry  & -1.93 & 1.31 & -0.03 & 0.75 \\
Employment in services & 2.02 & 1.86 & \textbf{2.08} & 0.83 \\
Lower education workers & -0.01 & 1.05 & -1.03 & 0.81 \\
Higher education workers & \textbf{3.86} & 0.66 & \textbf{4.14} & 0.78 \\
Regional knowledge capital & \textbf{0.08} & 0.03 & -0.02 & 0.05 \\
$\boldsymbol{W}$ Market size & \textbf{-2.04} & 0.18 & \textbf{-0.53} & 0.10 \\
$\boldsymbol{W}$ Population density & \textbf{-0.47} & 0.10 & \textbf{-0.35} & 0.06 \\
$\boldsymbol{W}$ Compensation per hour & -0.14 & 0.26 & \textbf{-0.63} & 0.27 \\
$\boldsymbol{W}$ Employment in industry  & -2.04 & 1.83 & 0.20 & 1.31 \\
$\boldsymbol{W}$ Employment in services & 1.47 & 1.57 & \textbf{-2.79} & 1.54 \\
$\boldsymbol{W}$ Lower education workers & 0.42 & 0.93 & -0.01 & 1.01 \\
$\boldsymbol{W}$ Higher education workers & \textbf{2.18} & 1.01 & \textbf{2.44} & 0.94 \\
$\boldsymbol{W}$ Regional knowledge capital & \textbf{-0.94} & 0.17 & \textbf{0.20} & 0.10 \\
\midrule
$\rho_o$, $\rho_d$ & \textbf{0.58} & 0.08 & \textbf{0.44} & 0.09 \\
$\phi_o^2$, $\phi_d^2$ & \textbf{0.70} & 0.08 & \textbf{1.28} & 0.13 \\
\midrule
Geographic distance & \textbf{-1.01} & 0.03 &    &  \\
Difference in tax rates & \textbf{1.30} & 0.63 &    &  \\
Common language & \textbf{0.51} & 0.07 &    &  \\
\bottomrule
\end{tabularx}%
\begin{tablenotes}
\item \textbf{Notes}: The model includes a constant. Results based on 15,000 Markov-chain Monte Carlo iterations, where the first 10,000 were discarded as burn-in. Estimates in bold are statistically significant under a 90\% confidence interval.
\end{tablenotes}
  \label{tab:emp_res}%
\end{threeparttable}%
\end{table}
\vspace{0.5cm}

From a theoretical point of view, we would also expect labour costs, measured in terms of \textit{compensation per hours}, to be an important determinant for attracting FDI inflows. This hypothesis is confirmed by inspecting the destination-specific results across all tables. Significant negative direct impacts of this variable can be observed throughout all stages of the value chain, both concerning the own region, as well as third regions. This corroborates the findings of \cite{ascani2016drives}, who study the location determinants of Italian multinational enterprises. \textit{Regional knowledge capital} as a pull-factor for upstream FDI inflows appears less relevant. Only the respective third-regional impact is significant, however, it appears comparatively muted.

Overall, the results for downstream FDI reported in Table \ref{tab:downstream} show a strong similarity to those of upstream FDI  (Table \ref{tab:upstream}). This resemblance can be observed for both origin- and destination-specific spatial determinants. For regions as a source of downstream FDI, Table \ref{tab:downstream} also highlights the key importance of agglomeration forces, proxied by the variables \textit{market size} and \textit{population density}. Both variables show a positive and significant direct impact for the generation of downstream FDI outflows, along with negative third-regional effects. These impacts, however, appear somewhat less pronounced as compared to upstream FDI. Similarly, the  impact of regional tertiary education attainment (\textit{higher education workers}) for downstream FDI outflows appears less accentuated as compared to upstream FDI outflows. As opposed to the results for origin-specific upstream FDIs, the third-regional effects of tertiary education attainment are insignificant. \textit{Regional knowledge capital} endowments, on the other hand, appear somewhat more important for generating downstream FDI as compared to upstream FDI, with positive direct, and negative third-regional effects. 

\begin{table}[h!tb] 
\centering
\begin{threeparttable}[b]
  \centering
  \caption{Posterior parameter estimates  for  FDI associated with downstream value chains. \label{tab:downstream}}
  \small
\begin{tabularx}{.9\linewidth}{@{}l c*{4}{Y}@{} }
\toprule
\multicolumn{1}{c}{\multirow{2}[2]{*}{Variable}} & \multicolumn{2}{c}{Origin} & \multicolumn{2}{c}{Destination} \\
   & \multicolumn{1}{c}{Mean} & \multicolumn{1}{c}{Std. Dev.} & \multicolumn{1}{c}{Mean} & \multicolumn{1}{c}{Std. Dev.} \\
\midrule
Market size & \textbf{0.65} & 0.10 & \textbf{1.26} & 0.06 \\
Population density & \textbf{0.24} & 0.08 & \textbf{0.17} & 0.05 \\
Compensation per hour & -0.43 & 0.33 & \textbf{-0.92} & 0.17 \\
Employment in industry  & -1.33 & 0.98 & \textbf{2.38} & 1.10 \\
Employment in services & 0.61 & 1.13 & \textbf{2.97} & 0.69 \\
Lower education workers & -1.03 & 0.74 & \textbf{-1.10} & 0.64 \\
Higher education workers & \textbf{2.90} & 0.64 & \textbf{3.29} & 0.89 \\
Regional knowledge capital & \textbf{0.38} & 0.05 & \textbf{-0.06} & 0.03 \\
$\boldsymbol{W}$ Market size & \textbf{-1.34} & 0.11 & \textbf{-0.55} & 0.15 \\
$\boldsymbol{W}$ Population density & \textbf{-0.63} & 0.19 & -0.13 & 0.09 \\
$\boldsymbol{W}$ Compensation per hour & \textbf{-0.58} & 0.21 & \textbf{-0.48} & 0.20 \\
$\boldsymbol{W}$ Employment in industry  & \textbf{-2.14} & 1.13 & 0.92 & 0.93 \\
$\boldsymbol{W}$ Employment in services & -1.15 & 1.68 & -1.31 & 0.98 \\
$\boldsymbol{W}$ Lower education workers & 1.01 & 1.01 & \textbf{1.53} & 0.68 \\
$\boldsymbol{W}$ Higher education workers & 1.78 & 1.04 & 1.24 & 0.94 \\
$\boldsymbol{W}$ Regional knowledge capital & \textbf{-0.92} & 0.17 & \textbf{0.26} & 0.10 \\
\midrule
$\rho_o$, $\rho_d$ & \textbf{0.42} & 0.12 & \textbf{0.52} & 0.08 \\
$\phi_o^2$, $\phi_d^2$ & \textbf{0.39} & 0.05 & \textbf{0.75} & 0.09 \\
\midrule
Geographic distance & \textbf{-0.85} & 0.03 &    &  \\
Difference in tax rates & \textbf{3.50} & 0.98 &    &  \\
Common language & \textbf{0.52} & 0.07 &    &  \\
\bottomrule
\end{tabularx}%
\begin{tablenotes}
\item \textbf{Notes}: The model includes a constant. Results based on 15,000 Markov-chain Monte Carlo iterations, where the first 10,000 were discarded as burn-in. Estimates in bold are statistically significant under a 90\% confidence interval.
\end{tablenotes}
  \label{tab:emp_res}%
\end{threeparttable}%
\end{table}
\vspace{0.5cm}

In line with the prevalent literature (see, among others, \citealt{leibrecht2014}, \citealt{laura2010evidence}, or \citealt{baltagi2007estimating}), the destination-specific regional determinants for downstream FDI also show a strong importance of the \textit{market size} and \textit{population density} variables as a means to attracting downstream-related FDI inflows. Similar to destination-specific upstream FDI, educational attainment (\textit{lower} and \textit{higher education workers}) and the \textit{compensation per hour} variable appear as important pull-factors. Concerning the regional industry mix, Table \ref{tab:downstream} suggests that higher shares in the industry and service sectors (\textit{employment in industry} and \textit{services}) appear to be significantly and positively associated with attracting downstream-related FDI inflows. An interesting result is given by a negative and statistically significant own-regional impact of the \textit{regional knowledge capital} variable. The estimated impacts, however, appear rather offset by the positive third-regional impacts. Similar results can also be found in work by \cite{dimitropoulou2013determinants}, a study on the location determinants of FDI for UK regions.

Empirical results for production-related FDI are summarized in Table \ref{tab:production}. Starting with the  origin-specific determinants of generating production FDI outflows, Table \ref{tab:production} shows not surprisingly a pronounced importance of regional \textit{market size} and \textit{population density}. Similar to the other types of FDI, both variables also exhibit significant negative third-regional effects. Interestingly, the source regional industry mix also appears to play a key role. Specifically, the \textit{employment in industry} variable shows a positive and highly significant direct impact of the origin region. The remaining origin-specific drivers are basically in line with those of the other types of FDI, most notably positive impacts of tertiary education attainment (\textit{higher education workers}) levels and \textit{regional knowledge capital} endowments. 

\begin{table}[h!tb] 
\centering
\begin{threeparttable}[b]
  \centering
  \caption{Posterior parameter estimates  for  FDI associated with production value chains. \label{tab:production}}
  \small
\begin{tabularx}{.9\linewidth}{@{}l c*{4}{Y}@{} }
\toprule
\multicolumn{1}{c}{\multirow{2}[2]{*}{Variable}} & \multicolumn{2}{c}{Origin} & \multicolumn{2}{c}{Destination} \\
   & \multicolumn{1}{c}{Mean} & \multicolumn{1}{c}{Std. Dev.} & \multicolumn{1}{c}{Mean} & \multicolumn{1}{c}{Std. Dev.} \\
\midrule
Market size & \textbf{1.01} & 0.18 & \textbf{0.94} & 0.09 \\
Population density & \textbf{0.14} & 0.07 & \textbf{-0.13} & 0.06 \\
Compensation per hour & 0.19 & 0.24 & \textbf{-1.21} & 0.14 \\
Employment in industry  & \textbf{2.78} & 0.80 & \textbf{4.04} & 0.73 \\
Employment in services & -0.07 & 0.98 & \textbf{3.10} & 0.52 \\
Lower education workers & 0.43 & 0.58 & -0.08 & 0.63 \\
Higher education workers & \textbf{2.68} & 0.68 & 0.52 & 0.62 \\
Regional knowledge capital & \textbf{0.20} & 0.06 & 0.00 & 0.04 \\
$\boldsymbol{W}$ Market size & \textbf{-1.11} & 0.15 & \textbf{-0.90} & 0.15 \\
$\boldsymbol{W}$ Population density & \textbf{-0.41} & 0.11 & 0.02 & 0.10 \\
$\boldsymbol{W}$ Compensation per hour & -0.38 & 0.29 & -0.61 & 0.50 \\
$\boldsymbol{W}$ Employment in industry  & 0.55 & 1.28 & -1.27 & 1.08 \\
$\boldsymbol{W}$ Employment in services & 0.13 & 1.40 & -0.85 & 0.81 \\
$\boldsymbol{W}$ Lower education workers & \textbf{1.04} & 0.44 & 1.14 & 0.92 \\
$\boldsymbol{W}$ Higher education workers & \textbf{2.25} & 0.83 & \textbf{2.03} & 0.79 \\
$\boldsymbol{W}$ Regional knowledge capital & \textbf{-1.21} & 0.14 & \textbf{0.25} & 0.11 \\
\midrule
$\rho_o$, $\rho_d$ & \textbf{0.77} & 0.05 & \textbf{0.32} & 0.11 \\
$\phi_o^2$, $\phi_d^2$ & \textbf{0.33} & 0.04 & \textbf{1.05} & 0.11 \\
\midrule
Geographic distance & \textbf{-0.96} & 0.03 &    &  \\
Difference in tax rates & \textbf{1.61} & 0.76 &    &  \\
Common language & \textbf{0.47} & 0.06 &    &  \\
\bottomrule
\end{tabularx}%
\begin{tablenotes}
\item \textbf{Notes}: The model includes a constant. Results based on 15,000 Markov-chain Monte Carlo iterations, where the first 10,000 were discarded as burn-in. Estimates in bold are statistically significant under a 90\% confidence interval.
\end{tablenotes}
  \label{tab:emp_res}%
\end{threeparttable}%
\end{table}
\vspace{0.5cm}

Inspection of the destination-specific determinants of production-related FDI, however, reveals markedly different patterns as compared to upstream and downstream FDI. Albeit the \textit{market size} shows a similar importance, along with negative third-regional effects, the direct impact of the \textit{population density} variable shows a negative and significant sign. Our estimation results thus show that production-oriented FDI activities are predominantly attracted by smaller regions in proximity to urban agglomerations. For upstream and downstream activities, however,  urban agglomerations seem to play a more central role. Moreover, our results imply that regional human capital endowments are particularly important for explaining upstream and downstream-oriented investment decisions. For production activities, the  importance of regional human capital endowments appears slightly less pronounced. These results corroborate the findings of \cite{strauss2009and}, and \cite{defever2006functional} by highlighting that industry-related location decisions typically focus on sectoral, rather than on functional aspects. The significant and positive own-regional, destination-specific industry mix (\textit{employment in industry} and \textit{services}) further underpins these findings. 

For attracting production-related FDI, Table \ref{tab:production} shows a particularly pronounced negative impact of the \textit{compensation per hour} variable of the host region. The negative direct impact on inflows is the strongest with a posterior mean of $-1.21$ for production-related activities. However, it is worth noting that the associated third-regional impacts on inflows are insignificant for production, whereas both downstream and upstream related FDI flows exhibit significant negative third-regional impacts. Our findings are moreover in line with \cite{fallon2014explaining} and \cite{crescenzi2013innovation}, who both find that locational drivers for production-related FDI flows differ from those associated with business service activities. 

\subsection*{Spatial-dependence and distance metrics}

This subsection discusses the results for the spatial autoregressive origin and destination random effects, as well as the estimates of intervening opportunities from the distance matrix $\boldsymbol{D}$. Inspection of  posterior estimates for the spatial latent random effects provides significant evidence for pronounced spatial dependence patterns in the random effects across all stages of the value chain. This finding holds true for both source- and host-regional heterogeneity in the sample. A comparison of their corresponding posterior means and standard deviations shows that all spatial autoregressive parameters are estimated with a high precision. The intensity of spatial dependence in the upstream- and downstream-specific latent unobservable effects appear similarly pronounced, with values ranging from $0.42$ to $0.58$. For production-related investment activities, the difference between $\rho_o$ and $\rho_d$ appears more pronounced, with the former being particularly sizeable ($0.77$), while the latter appears more muted.  

Rather similar results for upstream, downstream and production are also reported for the distance factors collected in matrix $\boldsymbol{D}$. As expected, the posterior mean estimates for \textit{geographical distance} are negative and significantly differ from zero for all types of investment activities. Moreover, the posterior standard deviations are comparatively small, indicating that the impact of \textit{geographic distance}  is estimated with a high precision. Higher geographic separation of two regions is thus associated with lower FDI activities, as increased distance often raises transportation, monitoring and thus investment costs. The negative impacts reported in Tables \ref{tab:upstream}, \ref{tab:downstream}, and \ref{tab:production} are in line with recent empirical results in FDI (\citealt{leibrecht2014}) and trade literature (\citealt{Krisztin2015}).

Our dummy variable measuring whether a pair of regions shares an official \textit{common language} proxies the cultural distance between regions in the sample. As expected, the reported posterior means show a positive sign and are significantly different from zero. The third distance variable in the matrix $\boldsymbol{D}$ measures the (country-specific) difference in \textit{corporate tax rates} between source and target regions. In line with theoretical and empirical literature on the location choice of multinationals, the tables report significant and positive impacts to regional FDI flows when corporate tax rates in the target region are lower than in the source region (see   \citealt{Bellak2009} and \citealt{strauss2009and}). The estimated posterior means for the difference in tax rates suggest that a $1\%$ decrease in the tax rate difference between source and destination regions results in a $1.3\%$ and $3.5\%$ increase in the number of FDI flows for downstream and upstream related activities, respectively.

\FloatBarrier

\section{Conclusions\label{s:7_conclusio}}
This paper presents an empirical study on the spatial determinants of bilateral FDI flows among European regions. Due to data scarcity on the subnational level, previous papers typically adopt a national perspective when analysing FDI dyads (see, for example, \citealt{leibrecht2014}). This paper thus provides a first spatial econometric analysis on the European regional level by explicitly accounting for origin-, destination-, and third region-specific factors in the analysis. The subnational perspective of our analysis allows us to study the spatial spillover mechanisms of regional FDI flows in more detail. Unlike recent  studies on the locational determinants of FDI inflows (see, for example, \citealt{ascani2016drives}, or \citealt{crescenzi2013innovation}), we model FDI decision determinants not only across destination regions but also across the origin regional dimension. Moreover, due to the well-known need to control for spatial dependence when modelling regional data \citep{LeSage2009}, we also capture spatial dependence through spatially structured random effects associated with origin and destination regions.

Our data comes from the \textit{fDi Markets} database, which contains detailed information on regional FDI activities using media sources and company data. The data from the \textit{fDi Markets} database also contains detailed sectoral information on the functional form of the FDI activity, which allows us to explicitly focus on FDI flows across different stages of the value chain. Specifically, the paper studies the origin- and destination-specific determinants of upstream, downstream, and production activities.

Our empirical results clearly indicate that both source and destination spatial dependence plays a key role for all investment activities under scrutiny. In line with recent literature, we find that regional market size, corporate tax rates, as well as third region effects appear to be of particular importance for all stages in the value chain. We moreover find that production-oriented FDI activities are predominantly attracted by smaller regions in proximity to urban agglomerations. For upstream and downstream activities, however, being in the same region as urban agglomerations seem to play a key role. Moreover, our results imply that regional human capital endowments are particularly important for explaining upstream and downstream-oriented investment decisions. For production activities, the  importance of regional human capital endowments are less accentuated. These results corroborate the findings of \cite{strauss2009and}, or \cite{defever2006functional} by highlighting that industry-related location decisions typically focus on sectoral, rather than on functional aspects. From an origin-specific perspective of FDI activities, our empirical results moreover clearly show that regional knowledge capital endowments appear crucial for host regions to produce FDI outflows. Similar to the results on the destination-specific factors for FDI inflows, we also find high education and agglomeration forces as particularly important aspects for host regional FDI outflows.

\section*{Declarations}
\textbf{Funding:} The research carried out in this paper was supported by funds of the Oesterreichische Nationalbank (project number: 18116), and of the Austrian Science Fund (FWF): ZK35. \\
\textbf{Conflict of interest:} The authors declare that they have no conflict of interest.\\
\textbf{Ethical approval:} This article does not contain any studies with human participants or animals performed by any of the authors.

\bibliographystyle{fischer}
\bibliography{library,MiscBib}

\newpage

\appendix

\renewcommand\thesection{\Alph{section}}
\setcounter{section}{1}

\section*{Appendix}
\renewcommand{\thetable}{A\arabic{table}}
\renewcommand{\thefigure}{A\arabic{figure}}
\setcounter{figure}{0}
\setcounter{table}{0}

\begin{table*}[htbp]
\caption{Classification of \textit{fDi Markets} business functions}\vspace*{-1.8em}
\footnotesize
\begin{center}
\begin{threeparttable}
\begin{tabularx}{\textwidth}{llG}
\toprule
\textbf{Classification} & \multicolumn{1}{c}{\textbf{Business activities}} & \multicolumn{1}{c}{\textbf{\% of classification}} \\
\midrule

\multicolumn{1}{l}{\multirow{6}[2]{*}{Upstream}} & Business Services & 64.0 \\
   & Design, Development and Testing & 10.8\\
   & Education and Training & 2.5 \\
   & Headquarters & 12.1 \\
   & \multicolumn{1}{l}{Information and Communication Technology and Internet Infrastructure} & 4.3 \\
   & Research and Development & 6.3 \\
   \midrule
\multicolumn{1}{l}{\multirow{6}[2]{*}{Downstream}} & Customer Contact Centre & 4.2 \\
   & \multicolumn{1}{l}{Logistics, Distribution and Transportation} & 26.9 \\
   & Maintenance and Servicing & 3.4 \\
   & Sales, Marketing and Support & 62.1 \\
   & Shared Services Centre & 2.0 \\
   & Technical Support Centre & 1.4 \\   
   \midrule
\multicolumn{1}{l}{\multirow{5}[2]{*}{Production}} & Construction & 21.0 \\
   & Electricity & 5.3\\
   & Extraction & 0.3 \\
   & Manufacturing & 72.1\\
   & Recycling & 1.3 \\

\bottomrule
\end{tabularx}%
\begin{tablenotes}[para,flushleft]
\footnotesize{\textit{Notes}: The last column indicates the percent of industry activities per FDI classification. The values are based on the total observed FDI flows in the \textit{fDi Markets} database targeting the selected NUTS-2 regions in the period 2003-2011.}
\end{tablenotes}
\end{threeparttable}
\end{center}
\label{tab:classification}
\end{table*}

\newpage

\setlength{\tabcolsep}{1pt}
\begin{threeparttable}[htbp]
  \centering
  \caption{List of regions in the study.}
  \tiny
\begin{tabularx}{\textwidth}{XXXXX}
    \toprule
\multicolumn{1}{c}{\textbf{Austria}} & \multicolumn{1}{c}{\textbf{France [continued]}} & \multicolumn{1}{c}{\textbf{Hungary}} & \multicolumn{1}{c}{\textbf{Poland [continued]}} & \multicolumn{1}{c}{\textbf{UK}}  \\
Burgenland (AT) & Languedoc-Roussillon & Dél-Alföld & Lódzkie & Bedfordshire and Hertfordshire \\
Kärnten & Limousin & Dél-Dunántúl & Lubelskie & Berkshire, Buckinghamshire and \\
Niederösterreich & Lorraine & Észak-Alföld & Lubuskie & \quad Oxfordshire \\
Oberösterreich & Midi-Pyrénées & Észak-Magyarország & Malopolskie & Cheshire \\
Salzburg & Nord - Pas-de-Calais & Közép-Dunántúl & Mazowieckie & Cornwall and Isles of Scilly \\
Steiermark & Pays de la Loire & Közép-Magyarország & Opolskie & Cumbria \\
Tirol & Picardie & Nyugat-Dunántúl & Podkarpackie & Derbyshire and Nottinghamshire \\
Vorarlberg & Poitou-Charentes & \multicolumn{1}{c}{\textbf{Ireland}} & Podlaskie & Devon \\
Wien & Provence-Alpes-Côte d'Azur & Border, Midland and Western & Pomorskie & Dorset and Somerset \\
\multicolumn{1}{c}{\textbf{Belgium}} & Rhône-Alpes & Southern and Eastern & Slaskie & East Anglia \\
Prov. Antwerpen & \multicolumn{1}{c}{\textbf{Germany}} & \multicolumn{1}{c}{\textbf{Italy}} & Swietokrzyskie & East Wales \\
Prov. Brabant Wallon & Arnsberg & Abruzzo & Warminsko-Mazurskie & East Yorkshire and \\
Prov. Hainaut & Berlin & Basilicata & Wielkopolskie & \quad Northern Lincolnshire \\
Prov. Liège & Brandenburg & Calabria & Zachodniopomorskie & Eastern Scotland \\
Prov. Limburg (BE) & Braunschweig & Campania & \multicolumn{1}{c}{\textbf{Portugal}} & Essex \\
Prov. Luxembourg (BE) & Bremen & Emilia-Romagna & Alentejo & Gloucestershire, Wiltshire and \\
Prov. Namur & Chemnitz & Friuli-Venezia Giulia & Algarve & \quad Bristol \\
Prov. Oost-Vlaanderen & Darmstadt & Lazio & Área Metropolitana de Lisboa & Greater Manchester \\
Prov. Vlaams-Brabant & Detmold & Liguria & Centro (PT) & Hampshire and Isle of Wight \\
Prov. West-Vlaanderen & Dresden & Lombardia & Norte & Herefordshire, Worcestershire \\
Région de Bruxelles-Capitale & Düsseldorf & Marche & \multicolumn{1}{c}{\textbf{Romania}} & \quad and Warwickshire \\
\multicolumn{1}{c}{\textbf{Bulgaria}} & Freiburg & Molise & Bucuresti - Ilfov & Highlands and Islands \\
Severen tsentralen & Gießen & Piemonte & Centru & Inner London \\
Severoiztochen & Hamburg & Provincia Autonoma di Bolzano/ & Nord-Est & Kent \\
Severozapaden & Hannover & \quad Bozen & Nord-Vest & Lancashire \\
Yugoiztochen & Karlsruhe & Provincia Autonoma di Trento & Sud - Muntenia & Leicestershire, Rutland and \\
Yugozapaden & Kassel & Puglia & Sud-Est & \quad Northamptonshire \\
Yuzhen tsentralen & Koblenz & Sardegna & Sud-Vest Oltenia & Lincolnshire \\
\multicolumn{1}{c}{\textbf{Czech Republic}} & Köln & Sicilia & Vest & Merseyside \\
Jihovýchod & Leipzig & Toscana & \multicolumn{1}{c}{\textbf{Slovakia}} & North Eastern Scotland \\
Jihozápad & Lüneburg & Umbria & Bratislavský kraj & North Yorkshire \\
Moravskoslezsko & Mecklenburg-Vorpommern & Valle d'Aosta/Vallée d'Aoste & Stredné Slovensko & Northern Ireland (UK) \\
Praha & Mittelfranken & Veneto & Východné Slovensko & Northumberland and Tyne and \\
Severovýchod & Münster & \multicolumn{1}{c}{\textbf{Latvia}} & Západné Slovensko & \quad Wear \\
Severozápad & Niederbayern & Latvija & \multicolumn{1}{c}{\textbf{Slovenia}} & Outer London \\
Strední Cechy & Oberbayern & \multicolumn{1}{c}{\textbf{Lithuania}} & Vzhodna Slovenija & Shropshire and Staffordshire \\
Strední Morava & Oberfranken & Lietuva & Zahodna Slovenija & South Western Scotland \\
\multicolumn{1}{c}{\textbf{Denmark}} & Oberpfalz & \multicolumn{1}{c}{\textbf{Luxemburg}} & \multicolumn{1}{c}{\textbf{Sweden}} & South Yorkshire \\
Hovedstaden & Rheinhessen-Pfalz & Luxemburg & Mellersta Norrland & Surrey, East and West Sussex \\
Midtjylland & Saarland & \multicolumn{1}{c}{\textbf{Netherlands}} & Norra Mellansverige & Tees Valley and Durham \\
Nordjylland & Sachsen-Anhalt & Drenthe & Östra Mellansverige & West Midlands \\
Sjælland & Schleswig-Holstein & Flevoland & Övre Norrland & West Wales and The Valleys \\
Syddanmark & Schwaben & Friesland (NL) & Småland med öarna & West Yorkshire \\
\multicolumn{1}{c}{\textbf{Estonia}} & Stuttgart & Gelderland & Stockholm & \\
Eesti & Thüringen & Groningen & Sydsverige & \\
\multicolumn{1}{c}{\textbf{Finland}} & Trier & Limburg (NL) & Västsverige & \\
Åland & Tübingen & Noord-Brabant & \multicolumn{1}{c}{\textbf{Spain}} & \\
Etelä-Suomi & Unterfranken & Noord-Holland & Andalucía & \\
Helsinki-Uusimaa & Weser-Ems & Overijssel & Aragón & \\
Länsi-Suomi & \multicolumn{1}{c}{\textbf{Greece}} & Utrecht & Cantabria & \\
Pohjois-ja Itä-Suomi & Anatoliki Makedonia, Thraki & Zeeland & Castilla y León & \\
\multicolumn{1}{c}{\textbf{France}} & Attiki & Zuid-Holland & Castilla-la Mancha & \\
Alsace & Dytiki Ellada & \multicolumn{1}{c}{\textbf{Norway}} & Cataluña & \\
Aquitaine & Dytiki Makedonia & Agder og Rogaland & Comunidad de Madrid & \\
Auvergne & Ionia Nisia & Hedmark og Oppland & Comunidad Foral de Navarra & \\
Basse-Normandie & Ipeiros & Nord-Norge & Comunidad Valenciana & \\
Bourgogne & Kentriki Makedonia & Oslo og Akershus & Extremadura & \\
Bretagne & Kriti & Sør-Østlandet & Galicia & \\
Centre (FR) & Notio Aigaio & Trøndelag & Illes Balears & \\
Champagne-Ardenne & Peloponnisos & Vestlandet & La Rioja & \\
Corsica & Sterea Ellada & \multicolumn{1}{c}{\textbf{Poland}} & País Vasco & \\
Franche-Comté & Thessalia & Dolnoslaskie & Principado de Asturias & \\
Haute-Normandie & Voreio Aigaio & Kujawsko-Pomorskie & Región de Murcia & \\
Île de France &  &  &    & \\
\bottomrule
\end{tabularx}%
  \label{tab:Appendix_regions}%
\end{threeparttable}%

\newpage

\subsection{Detailed description of the Bayesian Markov-chain Monte Carlo algorithm \label{s:Gibbs_sampler_detail}}



This section provides a detailed description of the employed Bayesian Markov-chain Monte Carlo (MCMC) algorithm. A similar version is employed by \cite{LeSage2007}, who use such a modelling strategy for estimation of knowledge spillovers (measured in terms of patenting dyads) in European regions. Specifically, their estimation approach relies on work by \cite{Fruhwirth-Schnatter2006}, who introduce a Bayesian auxiliary mixture sampling approach for non-Gaussian distributed data. This approach builds on a hierarchical data augmentation procedure by introducing $y_i+1$ latent variables for each observation $y_i$, where $y_i$ denotes the $i$-th element of $\boldsymbol{y}$ (with $i = 1,...,N$). 

In order to alleviate the implied computational burden, we rely on an improved version of this auxiliary mixture sampling algorithm \citep{Fruhwirth-Schnatter2013}. The algorithm tremendously reduces the number of latent parameters per observation. Specifically, the required number of latent parameters is reduced from $y_i+1$ to at most two per observation for Poisson distributed data \citep{Fruhwirth-Schnatter2013}. 

From a statistical point of view, $\lambda_i$ from Eq. (\ref{eq:random_effects_model}) can be interpreted as a parameter in a Poisson process describing occurring events in a given time interval, where $\lambda_i$ denotes the $i$-th element of the Poisson mean $\boldsymbol{\lambda}$. For illustration, imagine sorting all unique values of the observed FDI flows from lowest to highest. The Poisson process can be viewed as modeling -- given a specific number of FDI flows -- the probability of jumping from one unique value to the next. These two quantities can be characterized as so-called arrival and inter-arrival times. Motivated by this formulation, the distribution itself can be described using merely  arrival and inter-arrival times, derived from the rate of the process $\lambda_i$. The expected value of arrival time of $y_i$ is $\nicefrac{1}{\lambda_i}$ and it follows a Gamma distribution with shape one and rate equal to $y_i$. The inter-arrival times are by definition independent and arise from an exponential distribution with rate equal to $\lambda_i$. Based on this definition, we can model $\lambda_i$ if we sample from the inter-arrival time $\tau_{i1}$ between $y_i$ and $y_i + 1$, as well as for $y_i > 0$, the arrival $\tau_{i2}$ time for $y_i$. The main contribution of \cite{Fruhwirth-Schnatter2009} is that they introduce auxiliary variables for  $\tau_{i1}$ and  $\tau_{i2}$, conditional on $y_i$. 

For this purpose let us define the latent variables $\tau_{i1}$ and $\tau_{i2}$, based on the properties of arrival and inter-arrival times:
\begin{align}
\tau_{i1} &= \frac{\xi_{i1}}{\lambda_i}, \hspace{2em} \xi_{i1} \sim \mathcal{E}(1) \label{eq:tau1_prior}\\
\tau_{i2} &= \frac{\xi_{i2}}{\lambda_i}, \hspace{2em} \xi_{i1} \sim \mathcal{G}(y_i,1) \hspace{2em} \forall \,\, y_i > 0, \label{eq:tau2_prior}
\end{align}
where $\mathcal{E}(\cdot)$ denotes the exponential and $\mathcal{G}(\cdot,\cdot)$ the Gamma distribution. The arrival times $\tau_{i2}$ only apply for $y_i>0$, since zero values have by definition no arrival time. Eqs. (\ref{eq:tau1_prior}) and (\ref{eq:tau2_prior}) can be log-linearized in the following fashion:
\begin{align}
-\ln \tau_{i1} &= \ln \lambda_i + \varepsilon_{i1}, \hspace{2em} \varepsilon_{i1} = - \ln\xi_{i1} \label{eq:log_tau} \\
-\ln \tau_{i2} &= \ln \lambda_i + \varepsilon_{i2}, \hspace{2em} \varepsilon_{i2} = - \ln\xi_{i2} \hspace{2em} \forall \,\, y_i > 0,
\end{align}
where for $y_i = 0$ only Eq. (\ref{eq:log_tau}) is defined. Evidently, if  $\varepsilon_{i1}$ and $\varepsilon_{i2}$ would be Gaussian this would imply a linear model, which could  be easily sampled from. While $\varepsilon_{i1}$ and $\varepsilon_{i2}$ are not Gaussian per se, the distributions can be approximated by a mixture of Gaussians, from which sampling can easily be achieved \citep{Fruhwirth-Schnatter2009}.

In order to obtain a model which is conditionally Gaussian, the non-normal density can be approximated by a mixture of $Q(\nu)$ normal components, where $\nu$ denotes the shape parameter of a Gamma distribution. For sampling $\varepsilon_{i1}$ we can set $\nu = 1$, and in the case of sampling $\varepsilon_{i1}$ the rate $\nu$ would be equal to $y_i$. Therefore, the mixture of normal components can be generalised for both distributions. Thus, the mixture distribution is given by:
\begin{align}
p_{\varepsilon}(\varepsilon|\nu) \sim \sum_{q=1}^{Q(\nu)} w_q(\nu) \mathcal{N} \left[\varepsilon | m_q(\nu), s_q(\nu)\right],  \label{eq:Gaussian_mix}
\end{align}
where $w_q(\nu)$ denotes the weight, $m_q(\nu)$ the mean, and $s_q(\nu)$ the variance. These components, as well as $Q(\nu)$ directly depend on the choice of $\nu$. Values for all these parameters conditional on $\nu$ are provided in \cite{Fruhwirth-Schnatter2009}. To approximate the Poisson process through the Gaussian mixture in Eq. (\ref{eq:Gaussian_mix}), the additional latent discrete variable $\nu_{i1}$, and additionally in cases of $y_i>0$ the discrete variable $\nu_{i2}$ are introduced.

Given $\tau_{i1}$ and $\nu_{i1}$ and additionally for the case of $y_i>0$ $\tau_{i2}$ and $\nu_{i2}$, the conditional posterior of the Poisson model's slope parameters are Gaussian:
\begin{align}
-\ln \tau_{i1} &=  \ln \lambda_i + m(1) + \varepsilon_{i1}, &\hspace{2em} &\varepsilon_{i1}|\nu_{i1} \sim \mathcal{N}\left[0,s(1)\right]  \label{eq:normal_tau1} \\
-\ln \tau_{i2} &= \ln \lambda_i + m(\nu_{i2}) + \varepsilon_{i2}, 
&\hspace{2em} &\varepsilon_{i2}|\nu_{i2} \sim \mathcal{N}\left[0,s(\nu_{i2})\right] & \hspace{2em} \forall \,\, y_i > 0. \label{eq:normal_tau2}
\end{align}
We can easily sample from the distributions given in Eqs. (\ref{eq:normal_tau1}) and (\ref{eq:normal_tau2}) and therefore construct an efficient Gibbs sampling algorithm (for a detailed description, see Section \ref{s:Gibbs_sampler_detail} in the Appendix).

For Bayesian estimation, we have to define prior distributions for all parameters in the model. We follow the canonical approach and use a Gaussian prior setup for the parameters $\alpha_0$, $\boldsymbol{\beta}_o$, $\boldsymbol{\beta}_d$, $\boldsymbol{\gamma}_D$, $\boldsymbol{\delta}_o$, and $\boldsymbol{\delta}_d$ with zero mean and a relatively large prior variance of $10^4$. We follow \cite{LeSage2007} in our choice of priors for the spatially structured random effect vectors and set a normal prior structure $\boldsymbol{\theta}_o$ and $\boldsymbol{\theta}_d$, with with zero mean and $\phi_x^2 \left( \boldsymbol{A}_x \boldsymbol{A}_x \right)^{-1}$ variance, where $x \in [o,d]$ and $ \boldsymbol{A}_x = \boldsymbol{I}_n - \rho_x \boldsymbol{W}$. For the variance of the random effects $\phi_x^2$ we employ an inverse Gamma prior with rate equal to $5$ and the shape parameter to $0.05$. Following \cite{LeSage2007}, we elicit a non-informative uniform prior specification $\rho_x \sim \mathcal{U}(-1,1)$. 

\subsection{The Gibbs sampling scheme}

Let us collect the explanatory variables from Eq. (\ref{eq:random_effects_model}) in an $N \times P$ (with $P = 1 + 4 p_X + p_D$) matrix $\boldsymbol{Z} = [\boldsymbol{\iota}_N, \boldsymbol{X}_o, \boldsymbol{X}_d, \boldsymbol{D},\boldsymbol{W}_o\boldsymbol{X}_o,\boldsymbol{W}_d\boldsymbol{X}_d]$ and $\boldsymbol{\gamma} = [\alpha_0, \boldsymbol{\beta}_o',\boldsymbol{\beta}_d',\boldsymbol{\gamma}_D',\boldsymbol{\delta}_o',\boldsymbol{\delta}_d']'$. Thus, $\boldsymbol{\lambda} =  \exp\left( \boldsymbol{Z} \boldsymbol{\gamma} + \boldsymbol{V}_o\boldsymbol{\theta}_o + \boldsymbol{V}_d \boldsymbol{\theta}_d \right)$.

Moreover, let us denote the number of non-zero observations in $\boldsymbol{y}$ as $N_{y>0}$. Then, let $N_+ = N + N_{y>0}$ and let the $N_+ \times 1$ vector $\boldsymbol{y}_+$ be $\boldsymbol{y}_+ = [\boldsymbol{y}', \boldsymbol{y}_{y>0}']'$, where $\boldsymbol{y}_{y>0}$ contains all elements of $\boldsymbol{y}$ which are greater than zero. Moreover, let the $N_+ \times P$ matrix $\boldsymbol{Z}_+$ be $\boldsymbol{Z}_+ = [\boldsymbol{Z}', \boldsymbol{Z}_{y>0}']'$, where the matrix $\boldsymbol{Z}_{y>0}$ contains all rows of $\boldsymbol{Z}$ corresponding to $y_k>0$. In a similar fashion, we augment the dummy observation matrices $\boldsymbol{V}_o$ and $\boldsymbol{V}_d$ and denote the resulting $N_+ \times n$ matrices as $\boldsymbol{V}^+_o$ and $\boldsymbol{V}^+_d$. 

Accordingly we order the auxiliary variables corresponding to $\varepsilon_{i1}$ and $\varepsilon_{i2}$ and collect them into the following $N_+ \times 1$ auxiliary variable vectors as $\boldsymbol{\tau} = [\tau_{11} ,..., \tau_{N1},\tau_{12},...,\tau_{N_{y>0}2}]$ and $\boldsymbol{\nu} = [
\nu_{11}, ... , \nu_{N1} , \nu_{12} , ... , \nu_{N_{y>0}2}]$. Based on this, we define the $N_+ \times N_+$ variance matrix $\boldsymbol{\Omega}$. Additionally -- based on the definition of the Gaussian mixtures in Eqs. (\ref{eq:normal_tau1} - \ref{eq:normal_tau2}) -- an $N_+ \times 1$ vector of working responses $\tilde{\boldsymbol{y}}$ can be obtained conditional on  $\boldsymbol{\tau}$ and $\boldsymbol{\nu}$, so that  $\tilde{\boldsymbol{y}} = m(\boldsymbol{\nu}) - \ln \boldsymbol{\tau}$.

Given appropriate starting values the following Gibbs sampling algorithm can be devised:
\begin{enumerate}[I.]
\item Sample $\boldsymbol{\gamma}$ from its conditional Gaussian distribution $p(\boldsymbol{\gamma}|\cdot) \sim \mathcal{N}(\overline{\boldsymbol{\Sigma}}_{\boldsymbol{\gamma}}\overline{\boldsymbol{\mu}}_{\boldsymbol{\gamma}},\overline{\boldsymbol{\Sigma}}_{\boldsymbol{\gamma}})$, where
\begin{align*}
\overline{\boldsymbol{\Sigma}}_{\boldsymbol{\gamma}} = \left(\boldsymbol{Z}_+' \boldsymbol{\Omega}^{-1} \boldsymbol{Z}_+ + \underline{\boldsymbol{\Sigma}}_{\boldsymbol{\gamma}}^{-1} \right)^{-1} \hspace{2em} \overline{\boldsymbol{\mu}}_{\boldsymbol{\gamma}} = \boldsymbol{Z}_+' \boldsymbol{\Omega}^{-1} (\tilde{\boldsymbol{y}} - \boldsymbol{V}^+_o \boldsymbol{\theta}_o - \boldsymbol{V}^+_d \boldsymbol{\theta}_d  ) + \underline{\boldsymbol{\Sigma}}_{\boldsymbol{\gamma}}^{-1}\underline{\boldsymbol{\mu}}_{\boldsymbol{\gamma}}.
\end{align*}
$\underline{\boldsymbol{\Sigma}}_{\boldsymbol{\gamma}}$ denotes the $P \times P$ prior variance matrix and $\underline{\boldsymbol{\mu}}_{\boldsymbol{\gamma}}$ the $P \times 1$ matrix of prior means.
\item Sample $\boldsymbol{\theta}_x$ from their conditional distributions $p(\boldsymbol{\theta}_x|\cdot) \sim \mathcal{N}(\overline{\boldsymbol{\Sigma}}_{\boldsymbol{\theta}_x}\overline{\boldsymbol{\mu}}_{\boldsymbol{\theta}_x},\overline{\boldsymbol{\Sigma}}_{\boldsymbol{\theta}_x})$, where
\begin{align*}
\overline{\boldsymbol{\Sigma}}_{\boldsymbol{\theta}_x} = \left( \phi^{-2}_x {\boldsymbol{A}_x}'\boldsymbol{A}_x + \boldsymbol{V}_x^{+'} \boldsymbol{\Omega}^{-1} \boldsymbol{V}_x^+ \right)^{-1}
\hspace{2em} \overline{\boldsymbol{\mu}}_{\boldsymbol{\theta}_x} =  \boldsymbol{V}_x^{+'} \boldsymbol{\Omega}^{-1} \left( \tilde{\boldsymbol{y}} - \boldsymbol{Z}_+ \boldsymbol{\gamma} -  \boldsymbol{V}_x^+ \boldsymbol{\theta}_x \right).
\end{align*}
\item We sample $\phi^2_x$ from the conditional posterior, which is inverse Gamma distributed and given as $p(\phi^2_x|\cdot) \sim \mathcal{IG}(\overline{s}_x,\nicefrac{1}{\overline{v}_x})$ where:
\begin{align*}
\overline{s}_x = (n + \underline{s}_x)/2 \hspace{2em} \overline{v}_x = \left(\boldsymbol{\theta}_x' \boldsymbol{A}_x' \boldsymbol{A}_x  \boldsymbol{\theta}_x + \underline{s}_x\underline{v}_x \right)/2.
\end{align*}
$\underline{s}_x$ and $\underline{v}_x$ denote the prior rate and shape parameters of the inverse gamma distribution $\mathcal{IG}(\cdot,\cdot)$.

\item For $\rho_x$  the conditional posterior distribution is:
\begin{align*}
p(\rho_x|\cdot) \propto |\boldsymbol{A}_x|\exp\left( - \frac{1}{2\phi^2_x} \boldsymbol{\theta}_x' \boldsymbol{A}_x' \boldsymbol{A}_x  \boldsymbol{\theta}_x \right).
\end{align*}
Unfortunately, this is not a well-known distribution, thus -- as is standard in the spatial econometric literature -- we resort to a griddy Gibbs step (\citealt{ritter1992facilitating}) in order to sample from the conditional posterior for $\rho_x$.\footnote{An alternative, however, computationally more intensive approach also frequently used in the spatial econometric literature involves a Metropolis-Hastings step for the spatial autoregressive parameter (see, for example, \citealt{LeSage2009}).} For this purpose candidate values $\rho_x^*$ are sampled from $\rho_x^* = \mathcal{N}(\rho_x,\pi_{\rho_x})$, where $\pi_{\rho_x}$ is the proposal density variance, which is adaptively adjusted using the procedure from \cite{LeSage2009} and thus is constrained to a desired interval by the means of rejection sampling. The candidate values are evaluated using their full posterior distributions\footnote{In practice it is costly to evaluate the log-determinant directly. Instead we use an adapted version of the log-determinant approximation by \cite{Pace1997} for pre-calculation. }.

\item For $i=1,...,N$ we sample $\tau_{i1}$ from $\xi_{i1} \sim \text{Ex}(\boldsymbol{\lambda}_i)$ and set $\tau_{i1} = 1 + \xi_{i1}$. If $y_k>0$ then we additionally sample $\tau_{i2}$ from $\mathcal{B}(y_i,1)$ (where $\mathcal{B}(\cdot)$ denotes the Beta distribution) and set $\tau_{i1} = 1- \tau_{i2} + \xi_{i1}$.
\item For $i=1,...,N$ we sample $\nu_{i1}$ from the discrete distribution involving the mixture of normal distributions with $r=1,...,Q(1)$:
\begin{align*}
p(\nu_{i1} = r|\cdot) \propto w_r(1) \mathcal{N}\left[-\ln \tau_{i1}- \ln \lambda_i|  m_r(1), s_r(1)\right]
\end{align*}
and for $y_i>1$ we additionally sample $\nu_{i2}$ from the discrete distribution (with $r = 1,...,Q(y_i)$):
\begin{align*}
p(\nu_{i2} = r|\cdot) \propto w_r(y_i) \mathcal{N}\left[-\ln \tau_{i2}- \ln \lambda_i|  m_r(y_i), s_r(y_i)\right].
\end{align*}
With the sampled values for $\boldsymbol{\tau}$ and $\boldsymbol{\nu}$, we update $\tilde{\boldsymbol{y}} = \ln \boldsymbol{\tau} - {m}(\boldsymbol{\nu})$ and  $\boldsymbol{\Omega}$.
\end{enumerate}
This concludes the Gibbs sampling algorithm. The Markov-chain Monte Carlo algorithm cycles through steps I. to VI. $B$ times and excludes the initial $B_0$ draws as burn-ins. Inference regarding the parameters is subsequently conducted using the remaining $B-B_0$  draws.\footnote{Whether the MCMC algorithm converged can be easily verified using convergence diagnostics  by \cite{geweke1991evaluating} or \cite{raftery1991many}. For the present application we utilised an implementation of these convergence diagnostics from the R \textbf{coda} package.}

\end{document}